# Long-exposure Camera Readout for Coherent Control of Nitrogen-Vacancy Center Spins in Diamond

Jiahui Chen[1†], Qilong Wu[1†], Huihui Yu[1], Yi-Dan Qu[1], Yuan Zhang[1,2*], Xun Yang[1*], Chong-Xin Shan[1*]

1. Henan Key Laboratory of Diamond Optoelectronic Materials and Devices, Key Laboratory of Material Physics Ministry of Education, School of Physics and Microelectronics, Zhengzhou University, Zhengzhou 450052, China
2. Institute of Quantum Materials and Physics, Henan Academy of Sciences, Zhengzhou 450046, China

[†]The authors contribute equally to the work.
*Contract emails: yzhuaudipc@zzu.edu.cn; yangxun9103@zzu.edu.cn; cxshan@zzu.edu.cn

Camera-based wide-field quantum noise spectroscopy (QNS) with nitrogen-vacancy (NV) center spins in diamond can be used to extract DC and AC magnetic field noise with sub-micrometer spatial resolution, but its realization is currently limited by the conflict between the optimal optical spin readout time (normally below one microsecond) and the minimal camera exposure time (normally tens of microsecond). In the present article, we demonstrate fully coherent control of NV center spins via Rabi oscillation, Ramsey, Hahn echo and relaxometry experiments, i.e. the core of QNS, with a home-built camera-based setup, and achieve an unexpectedly high contrast of 12% with an optical spin readout time of 500 microseconds. We explain the absence of the above conflict in our system with the optical spin readout mechanism under weak laser illumination, where spin-lattice relaxation and weak optical pumping result in a rather slow reduction of contrast with increasing integration time. The revealed mechanism is instructive for the construction of the wide-field QNS with a sCMOS or EMCCD camera, and its application in the studies of magnetic material or superconducting material.

## I. INTRODUCTION

Nitrogen-vacancy (NV) center in diamond, formed by a substitutional nitrogen atom adjacent to a vacancy in carbon atom lattice, features rich electronic and spin degree of freedom, which can be explored and controlled with optical and microwave radiation [1]. The combination of spin-preserved radiative channel and spin-sensitive inter-system crossings enables the convenient spin initialization and readout with optical means. In addition, NV center spin has a relatively long coherence time at room temperature due to a rather clean spin environment, and can couple with multiple physical fields, such as magnetic field, temperature field, electric field and stress field [1]-[8]. All these features promote NV center as an ideal system in the field of quantum computing, quantum information, and quantum sensing [9]-[13].

Quantum noise spectroscopy (QNS) of NV center utilizes the characteristic times $T_1$, $T_2$, and $T_2^*$ of NV center spin to detect magnetic noise, current noise, and other noise within or on the surface of materials. The NV spin depolarization time $T_1$ is sensitive to high-frequency (~GHz) magnetic noise, the decoherence time $T_2$ is mainly influenced by mid-frequency (~kHz–MHz) magnetic noise, whose response window can be also tuned with advanced quantum protocols (such as CPMG or XY8), and the free induction decay time $T_2^*$ is highly sensitive to static or quasi-static magnetic field distributions or local structural inhomogeneities. So far, QNS has been widely applied to investigate many-body physics in diverse condensed matter systems [14], such as identification of anomalous noise in graphene arising from electron-phonon Čerenkov instabilities [15]; visualization of ferromagnetic–antiferromagnetic phase transitions within Moiré superlattices formed by twisted van der Waals materials [16]; determination of static and dynamic critical exponents in high-temperature superconductors [17]-[18].

In the QNS, it is essential to place the NV center probe close to the explored material ranging from nanometer to micrometer range, and simultaneously realize the coherent control of NV spins to extract the characteristic times $T_1$, $T_2$, and $T_2^*$. In the actual realization of QNS, a single photon detector is usually used to detect weak fluorescence of NV centers, and is often combined with scanning microscopic technique [18] or scanning probe technique to acquire spatial information [19]. However, the coherent control of NV centers usually takes quite long time, and thus leads to great challenge in maintaining the stability of the probing system, and relatively long time for the whole experiments (usually several hours). To solve this problem, one can use a camera to detect the NV fluorescence from shallow NV centers near the diamond surface, and realize wide-field QNS in parallel. Although wide-field magnetic field imaging has been routinely demonstrated using a quantum diamond microscope [20]-[22], wide-field QNS has not yet been realized. This is due to the conflict between the optimal optical readout time (less than one microsecond) and the minimal camera exposure time (usually tens of microseconds).

In the present article, we demonstrate fully coherent control of NV center spins via Rabi oscillations, Ramsey, Hahn echo, and relaxometry experiments using a camera-based integrated setup, thus realizing the core techniques of wide-field QNS. Importantly, we achieve an unexpectedly high contrast of 12% with 500 μs long camera exposure time, far beyond the minimal exposure time of 10 μs, and explain the absence of the above conflict with the unexplored optical spin readout mechanism under weak laser illumination, where spin-lattice relaxation and weak optical pumping result in a rather slow reduction of contrast with increasing integration time. The revealed mechanism is instructive for the construction of wide-field QNS with a sCMOS camera, and its application in the studies of magnetic material, superconducting material and other condensed matter systems with sub-micrometer resolution.

## II. NV Center and Experimental Setup

The NV center is formed by a substitutional nitrogen atom and an adjacent vacancy in the carbon lattice [Fig. 1(a)]. Because each carbon atom can form bonds with four neighboring atoms, the vacancy can take four possible positions relative to the nitrogen atom, leading to four orientations. NV center can be either in a neutral form (NV$^0$) or a negatively charged form (NV$^-$). Since only the latter is relevant for the quantum technology, we will consider exclusively it in the following. NV center has unique electronic and spin energy level structure, as shown in Fig. 1(b), which includes a triplet ground state $^3A_2$, a triplet excited state $^3E$ and two singlet states $^1A_1$, $^1E$. Each of the triplet states consists of three spin level, which can be labeled by three projection numbers $m_s = 0, \pm 1$ of magnetic field along the NV axis. The $m_s = \pm 1$ levels are degenerated in energy, and they are approximately 2.87 GHz, and 1.42 GHz above the $m_s = 0$ level for the $^3A_2$ and $^3E$ state.

NV center can experience rich processes as shown in Fig. 1(b). It can be optically excited from the $^3A_2$ state to the $^3E$ state (green arrow). The excited NV can decay directly to the $^3A_2$ ground state via spontaneous emission (red arrows) or through non-radiative and intersystem crossing via the $^1A_1$, $^1E$ level (dashed arrows). Since the intersystem crossing is faster for the $^3E$, $m_s = \pm 1$ spin level than that for the $^3E$, $m_s = 0$ spin level, a continuous excitation of the NV center leads to an effective population transfer from the $m_s = \pm 1$ to the $m_s = 0$ spin level, i.e. optically-induced spin polarization. This process can compensate the spin-lattice relaxation (dotted arrows), induced by the stimulated Raman scattering of lattice phonon [23], and lead to a population on the $m_s = 0$ spin level around 80% at the optimal condition. Because of the imbalanced intersystem crossing, the excited NV center on the $^3A_2$, $m_s = 0$ spin level emits more photons than that on the $^3A_2$ $m_s = \pm 1$ spin level, leading to the optical readout of the NV spin state.

Figure 1(c) shows the schematic of our experimental setup. We build an electronic box based on microcontroller, microwave source chips and a digital delay / pulse generator. The computer commands the electronic box to send pulse sequences to control the on/off states of the microwave, camera, and acoustic-optical modulator (AOM). A microwave radiation around 2.87 GHz from a chip-based source is delivered to a copper microwire on the top of diamond sample after being attenuated, switched and amplified. A laser beam is delivered to the diamond sample after being switched with AOM and focused on the diamond sample through an objective. A radiofrequency radiation of 100 MHz from another chip-based source is used to power the AOM after the similar operations. An industrial camera is used to capture a series of fluorescence images of the sample. More details of the setup are presented in the Sec. S1 of the Supplementary Materials (SM).

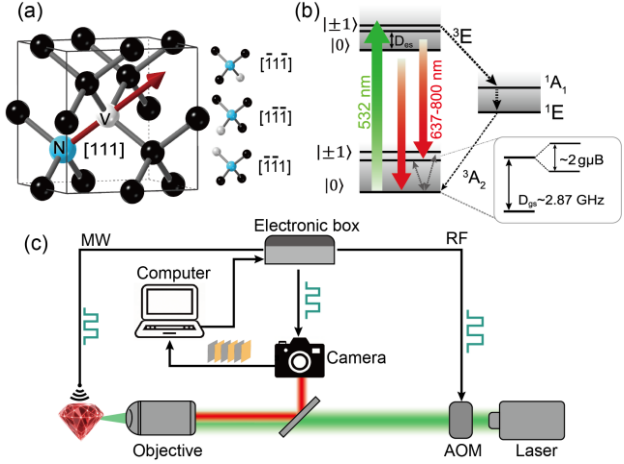

**FIG 1. NV center structure and experimental setup.** (a) Atomic structure of NV center consisting of a nitrogen atom (blue sphere) and a vacancy (white sphere) in carbon atom lattice (black spheres), and four possible orientations of the NV center. (b) Energy level diagram of the negatively charged NV⁻ center. (c) Schematic of the experimental setup, where the electronic box controls the pulse sequence of the microwave, the laser (through acoustic optical modulator, AOM) and the camera image acquisition, and the computer processes a series of images to obtain the expected results.

## III. Experimental and Theoretical Results

In this section, we demonstrate the experiments on the coherent control of NV center spin with camera-based integrated setup. Before presenting the results, we briefly describe the sample preparation. We grow a diamond sample with microwave plasma chemical vapor deposition method with 0.5 sccm of 5% nitrogen in our lab, and irradiate the sample with electrons with 10 MeV energy and $5 \times 10^{17}$ electrons/cm² dose, and anneal it at a temperature of 1000 ºC for 2 hours. The primary characterization with a commercial setup from CIQTEK indicates that this sample has a NV⁻ concentration of 0.4 ppm, the characteristic time $T_2^* = 560$ ns, $T_2 = 18.6$ μs and $T_1 = 5.1$ ms. Based on the literature [24], we estimate the nitrogen concentration of [N]=8.6 ppm from the $T_2$ characteristic time, leading to a conversion ratio [NV⁻]/[N] =4.65%. In the experiments, we glue the diamond sample on a printed circuit board, and then solder a copper microwire of 15 μm diameter across the sample surface. The wire is soaked with an ethanol solution, and is firmly attached to the diamond surface after the solution evaporation, which can increase the coupling of the microwave radiation to the NV centers near the surface. Subsequently, the sample holder is attached to a 3D positioning stage, and the laser beam is focused on the NV centers near the microwire.

### A. Coherent Control of NV Center Spin

In our experiment, the fluorescence imaging of the sample is taken for a given exposure time of 500 μs under laser illumination with a spot diameter of 100 μm and a power of 6 mW [inset of Fig. 2(a)]. We explore the region-of-interest feature of the camera to accelerate the image capture, and consider the area with fluorescence larger than 85% of the maximal value. Then, we generate a pulse sequence to switch on and off the microwave radiation of 24.6 dBm, and record the averaged fluorescence of the chosen area as the signal $I_s$ and the reference $I_r$ [upper part of Fig. 2(a)]. By repeating the experiment for different microwave frequency $f$, we obtain the optically detected magnetic resonance (ODMR) spectrum with a dip around 2.87 GHz [Fig. 2(a)]. By fitting the curve with single Lorentzian function, $C(f) = C_1 (\gamma/2)^2/[(f - f_0)^2 + (\gamma/2)^2] + C_2$, we obtain the maximal contrast $C_1 \approx -13.6\%$, the center frequency $f_0 = 2869.8$ MHz, and the linewidth $\gamma = 17.7$ MHz. When applying a magnetic field by moving a magnet over the sample with a 3D displacement stage, the ODMR spectrum becomes split into multiple dips. Depending on the relative orientation of the applied magnetic field with respect to the NV center axes [right insets of Fig. 2(b)], we can obtain the ODMR spectrum with two, four, six and eight dips. In particular, in the spectrum with four dips, the leftmost and rightmost dip can be attributed to the $m_s = 0 \rightarrow -1$ and $m_s = 0 \rightarrow +1$ transition of the NV centers, whose axis is parallel to the applied magnetic field, and the inner dips are caused by the transitions of other NV centers, whose axis is about 109.47º with respect to the magnetic field. By fitting the four dips with Lorentzian function, we extract the center frequency, the linewidth and the maximal contrast, and summarize them in the Tab. S1 of the SM. The frequencies of outmost dips are 2759.08 MHz and 2981.12 MHz, respectively, and their difference $\delta f$ =222.04 MHz can be used to refer the strength $B = \delta f/\gamma$ =39.65 Gauss (with the gyromagnetic ratio $\gamma = 2.8$ MHz/Gauss) of the applied magnetic field. In the following experiments, we assume that the microwave is resonant with the leftmost dip, and the spin levels $m_s = 0, -1$ are explored.

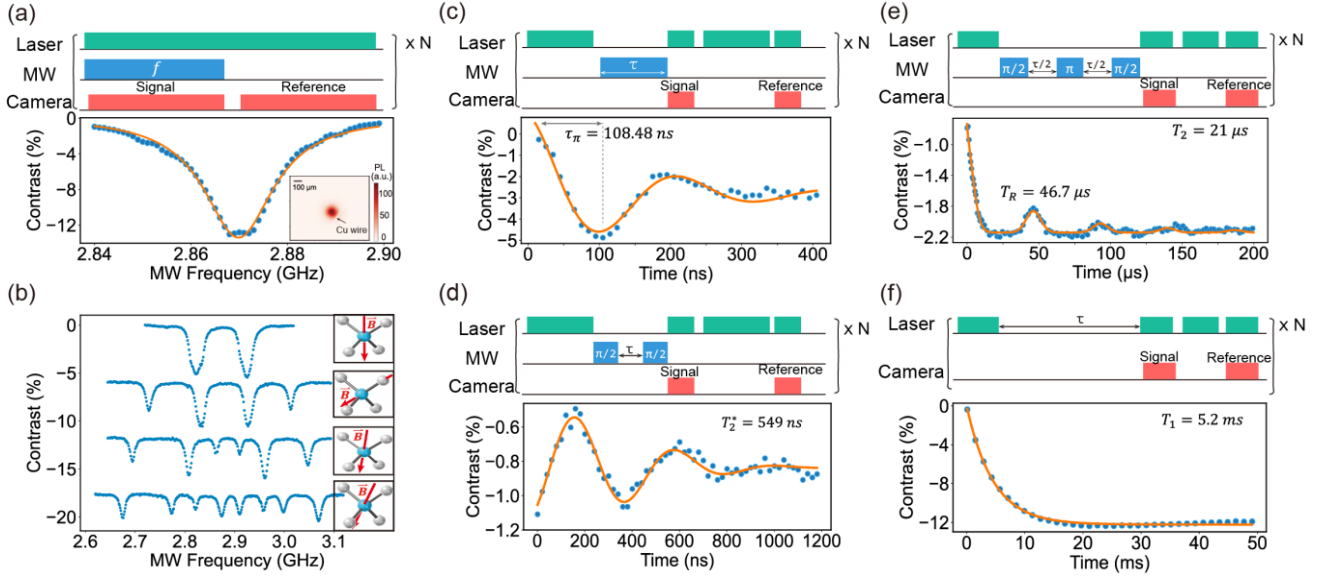

**FIG. 2. Coherent control of NV center spin.** Upper parts of the panels (a, c-f) show the pulse sequences of laser, microwave and camera for ODMR (a), Rabi oscillation (c), Ramsey (d), Hahn echo (e) and spin relaxometry (f) experiments, respectively. Lower parts of these panels show the obtained contrast (blue dots) as a function of the microwave frequency (a,b), the microwave pulse length (c) or the free evolution time (d-f), and the fittings (orange curve) with the expressions stated in the main text. The inset of the panel (a) shows the fluorescence imaging. The microwave radiation has a power around 24.6 dBm, and a frequency of 2759.08 MHz for the experiments (c-f). The focused laser has a power of 6 mW, the laser pulse for the NV spin initialization is 5 ms long. The camera exposure time is 500 μs. For more details, see the text.

In the Rabi oscillation experiment, we apply a pulse sequence shown on the top of Fig. 2(c), which includes a laser pulse of 5 ms long to initialize the NV center ensemble, and a microwave pulse of varying length $\tau$ that manipulates the quantum state of the system, and finally a readout pulse of 500 μs to trigger the camera to acquire the average fluorescence of chosen region, which is denoted as a signal. To eliminate slowly varying environment noise, the sequence also includes the second part without the microwave and the averaged fluorescence is denoted as the reference. When repeating the sequence five times, we obtain the contrast as a function of the microwave pulse length $\tau$, which shows an oscillatory and damped behavior [lower part of Fig. 2(c)]. By fitting the result with a damped cosine function $C(\tau) = C_0 \cos(\omega\tau + \varphi) \exp(-\tau/\tau_R) + C_1$, we obtain the amplitude $C_0 = -0.0377$, the oscillation frequency $\omega = 28.96$ MHz, the initial phase $\varphi = -3.1$ rad, the decay time $\tau_R = 138.68$ ns and the maximal contrast $C_1 = -2.8\%$. Then, we can determine the length of the $\pi/2-$ and $\pi-$ microwave pulses as $\tau_{\pi/2} = \pi/(2\omega) = 54.24$ ns, $\tau_\pi = \pi/\omega = 108.48$ ns, which will be used for the latter experiments.

In the Ramsey experiment, we apply the pulse sequence shown on the top of Fig. 2(d), which is similar to the previous one except that it replaces the single microwave pulse with two $\pi/2$ microwave pulses separated by a free evolution period of varying length $\tau$. After running experiments five times and calculating the average, we obtain the contrast as a function of $\tau$, which also shows an oscillatory behavior with damped amplitude [lower part of Fig. 2(d)]. By fitting the curve with the function $C(\tau) = C_0 \cos(\omega\tau + \varphi) \exp[-(\tau/T_2^*)^2] + C_1$. We obtain the amplitude $C_0 = 3.17 \times 10^{-3}$, the frequency $\omega = 14.5$ MHz, the initial phase $\varphi = -8.62$ rad, the decoherence time $T_2^* = 549$ ns, and the steady-state contrast $C_1 = -0.84\%$.

In the Hahn echo experiment, we apply the pulse sequence shown on the top of Fig. 2(e). By averaging over ten experiments, we obtain the contrast as a function of the total free evolution time $\tau$, which shows a fast decay for short time and several revived peaks with reduced amplitude at later time [lower part of Fig. 2(e)]. The revived peaks are caused by the hyperfine interaction with nearby $^{13}C$ nuclear spins under the magnetic field B = 39.65 Gauss, and reflects the revived coherence at the nuclear spin precision frequency. By fitting the curve with the expression $C(\tau) = C_0 \exp[-(\tau/T_2)^p] \sum_i \exp[-(\tau - i*\tau_R)^2/\tau_W^2] + C_1$, we obtain the amplitude $C_0 = -0.014$, the dephasing time $T_2 = 21$ μs, the stretch factor $p = 0.61$, the period $\tau_R = 46.7$ μs and width $\tau_W = 7.2$ μs of the revived peaks, and the contrast $C_1 = -2.14\%$. These results are consistent with other reported data as well as the results of theoretical simulations [25]-[26].

In the spin relaxometry experiment, we apply the pulse sequences shown on the top of Fig. 2(f). By carrying out the experiments five times and averaging the results, we obtain the contrast as a function of τ as shown in Fig. 2 (f), which shows a decayed behavior. By fitting the curve with an exponential function $C(\tau) = C_0 \exp(-\tau/T_1) + C_1$, we obtain the amplitude $C_0 = -0.015$, the relaxation time $T_1 = 5.2$ ms, and the maximal contrast $C_1 = -12.2\%$.

## B. Theoretical Exploration of Optical Spin Readout Mechanism

By analyzing the above experiments, we realize that the maximal contrast $C_{T_1}^{max}$ in the spin relaxometry experiment (about 12%) is achieved when the NV centers evolve to the thermal equilibrium state. Since the NV centers are also prepared in such a state during the NV spin initialization phase, $C_{T_1}^{max}$ can be obtained by using the fluorescence in this phase and the readout phase for zero delay time $\tau = 0$ as the signal and reference, respectively. Considering the relatively long time for the NV spin initialization (about 5 ms) and readout (about 500 μs), it is surprising that the coherent control of NV center spin is feasible with our setup, and a relatively large contrast has been achieved.

In the majority of experiments, single photon detectors are used to record the NV fluorescence, and the laser pulses for the NV spin initialization and readout are normally limited to few microseconds and hundreds of nanoseconds [1], respectively. To resolve the apparent discrepancy, we have estimated the laser power density in our study and earlier studies. In the usual experiments with single NV center, the laser spot has a diffraction-limited area of about 0.22 μm² and a power around hundreds of μs [27]. Assuming the laser power of 0.33 mW, we estimate a laser power density 1.5 mW/μm². In our experiments, the laser spot has an area of about $7.85 \times 10^3$ μm² and a power of 6 mW [inset of Fig. 2(a)], and thus the laser power density is estimated as 0.76 µW/µm$^2$, which is about three orders of magnitude weaker. Thus, to resolve the aforementioned discrepancy, we would like to compare theoretically the dynamics of the NV center during the different phase of the experiments for a strong and weak laser pumping in Fig. 3.

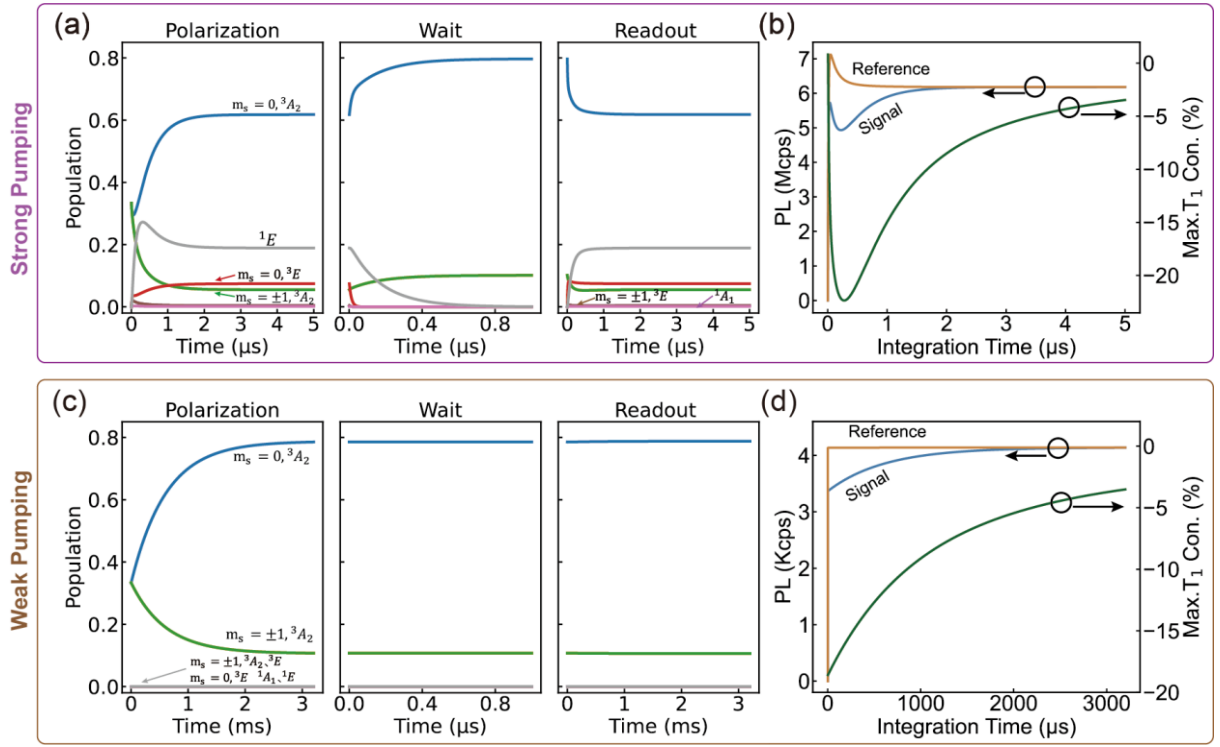

**FIG. 3. Comparison of the optical spin readout mechanism under strong (a-b) and weak (c-d) laser illumination.** In the panels (a) and (c), the left, middle and right part show the population dynamics of various levels during the phase of the NV spin initialization, the free evolution and the readout, respectively. In the panels (b) and (d), the fluorescence during the first and last phase is defined as the signal and reference, and the contrast is defined as the ratio of their difference to the reference for a given integration time. Note that this contrast corresponds to the maximal contrast in the spin relaxometry experiment. The optical pumping rate is assumed as 10 MHz and 5 kHz for the strong and weak laser illumination, respectively.

In the theoretical study, we model the NV center as an eight levels system, establish a quantum master equation to describe the system dynamics, and utilize the parameters for various processes reported in the literature [28]-[29], see Sec. S3 of SM for more details. In the simulations shown in Fig. 3(a,b) and (c,d), the optical pumping rate is assumed as 10 MHz and 5 kHz for a laser power of 0.33 mW and 0.16 µW, respectively, which correspond to a laser power density of 1.5 mW/µm$^2$ and 0.72 µW/µm$^2$ for a diffraction-limited illumination, respectively. Since these values are close to the values estimated above, the simulations are justified to resolve the discrepancy. Note that the laser power of 0.16 µW is rarely used in the single NV experiments due to the extremely weak fluorescence, and thus the readout mechanism under this condition has not been studied yet, as far as we know.

We present firstly the NV center dynamics under the strong optical pumping [Fig. 3(a)]. In the phase of NV spin initialization (left part), the population on the $^3A_2$, $m_s$=0 level increases and approaches a constant around 0.6 in less than 3 µs after dropping slightly in about 0.5 µs, and the population on the $^3A_2$, $m_s$=±1 levels decreases and approaches a constant around 0.05. At the same time, the population on the $^3E$, $m_s$=0 and $m_s$=±1 levels behave similarly with slightly less value, indicating the significant optical pumping, and the population of the $^1E$ level approaches a constant about 0.2 after reaching a local maximum in about 0.25 µs, suggesting a strong population trapping. To remove the population trapping effect, we introduce a waiting period without laser illumination about 1 µs to let the population decay to the $^3A_2$, $m_s$=0 and $m_s$=±1 levels (middle part). In the phase of the NV spin readout (right part), the population of the $^3A_2$, $m_s = 0, \pm 1$ levels decrease and approach constant values in about 1 µs, while those of the $^3E$, $m_s = 0, \pm 1$ levels and the singlet $^1E$ level increase and then reach the steady-state. Figure 3(b) shows that the NV fluorescence decays slowly to a constant value of about 6.18 Mcps in about 2 µs after reaching a maximum of 7.12 Mcps in about 46 ns in the spin readout phase, and it approaches the same constant value after reaching a local minimum about 4.93 Mcps in about 220 ns in the spin initialization phase. By defining these fluorescences as the reference and signal and then integrating them for given time window, we obtain the maximal contrast of the spin relaxometry experiment as a function of the integration time [green curve and right axis of Fig. 3(b)]. We find that the contrast decreases sharply to a value around -22.4% for the integration time of 0.25 µs, and then the absolute value decays exponentially with a characteristic time of 1.26 µs.

We analyze the NV center dynamics under the weak laser illumination [Fig. 3(c,d)]. Because the optical pumping rate of 5 kHz is three orders of magnitude smaller than the spontaneous emission rate of more than 75 MHz of the $^3E$ $m_s$=0,±1 levels, the population on these levels can decay immediately back to the $^3A_2$ $m_s$=0,±1 levels. Thus, we can eliminate adiabatically the excited levels, and obtain effective population transfer rates between the three ground levels, see Sec. S4 of the SM. Because the decay rate from the $^3E$, $m_s$=±1 level to the $^1A_1$ level is faster than from the $^3E$, $m_s$=0 level, the effective cooling rate of the $^3A_2$, $m_s$=±1 levels is much larger than the pumping rate of the $^3A_2$, $m_s$=0 level. As a result, the population of the former levels is transferred to the latter level, which competes with the spin-lattice relaxation and makes the population reach steady-state values [Fig. 3(c)]. Surprisingly, the populations of the $^3A_2$, $m_s$=0, ±1 levels for the weak optical pumping reach similar values as achieved for the strong optical pumping. Following the population dynamics, the NV fluorescence in the readout phase remains as a constant after jumping rapidly, while that in the NV spin initialization phase approaches gradually this constant value after jumping to a smaller value. As a result, the maximal contrast in the spin relaxometry experiment jumps quickly to the minimal value of -18.5% in about 5 µs, and then the absolute value decays slowly with a characteristic time of 1.12 ms [Fig. 3(d)].

Comparing Fig. 3 (a,b) with Fig. 3 (c,d), we see that the decay from the NV excited levels and the spin-lattice relaxation play an essential role in the optical spin readout under the strong and weak optical pumping, respectively, leading to small and larger integration time windows for significant contrast. In the latter case, for the integration time of 500 μs, the maximal contrast reaches about -12%, and is consistent with the experiment result shown in Fig. 2(f). Thus, the experiments demonstrated in Fig. 2 rely actually on the optical spin readout mechanism under the weak optical pumping.

### IV. Discussion and Conclusion

In the following, we analyze the limitations of our current setup, and outline the routes to reach the wide-field QNS experiments. To this end, we have firstly prepare a diamond sample with shallow NV centers by irradiating a CVD diamond sample grown in our lab with nitrogen ions of an energy 10 MeV and a dose $5 \times 10^{17}$ /cm$^2$, and then annealing the sample at temperature 1000 $^o$C for 2 hours. The preliminary characterization with a commercial setup indicates that the NV center layer has a surface density $4.95 \times 10^4$ /μm$^2$. Based on the reports in the literature [30], we estimate the layer thicknesses around 50 nm. The NV centers in such a sample can be used to detect DC and AC magnetic field noise in MHz and GHz from condensed matter samples [18], such as superconducting material, providing insights into carrier or spin dynamics involved.

We have carried out the experiment with the above sample, and found that an exposure time about 50 ms is required to observe the fluorescence under the laser illumination with a radius of r = 50 μm and power 6 mW. This occurs since the number of NV centers $3.88 \times 10^8$ is about two orders of magnitude smaller than $9.3 \times 10^{10}$ for the sample considered above. Here, the latter number is estimated by considering the NV concentration [NV$^-$]= 0.4 ppm = 0.704 $\times 10^5$ μm$^{-3}$ and modeling the laser spot as a sphere with a volume V= $4\pi r^3/3$ = $2.62 \times 10^5$ μm$^3$. However, according to Fig. 3(d), we estimate that the contrast approaches almost zero. To solve this problem, we can adopt other pulse schemes [21] to cover repeated pulse sequences within single camera exposure, but the total length of the experiment can become longer than the maximal camera exposure time about 2 s. If one increases the laser power by 100 times, the exposure time can be reduced to 500 μs as considered so far, but the decay of the contrast as shown in Fig. 3(d) becomes also reduced roughly by two orders of magnitude, and is beyond the minimal camera exposure time of 10 μs.

To solve the conflict between the camera exposure time and the integration window to reach acceptable contrast, it is necessary to replace our industrial camera with high sensitivity sCMOS or EMCCD camera with single photon detection ability. To assess this possibility, we have expanded the current experimental setup by introducing a sCOMS camera (Dhyana 400BSI V3) and utilizing a higher-power laser (MGL-S-532-300 mW). An optical beam expander with 5 magnification factor is positioned in front of the laser to broaden the beam. Subsequently, a 50 mm lens is employed to focus the beam onto the back focal plane of the objective lens. The resulting image is then projected onto the sCOMS camera via a lens with 200 mm focal length, thereby achieving an imaging area with a diameter of 180 μm and an exposure time of 250 μs. The laser power at the output of the objective lens is measured as 130 mW, and the laser density is estimated as 5.1 μW/μm$^2$. By considering the corresponding optical pumping rate 34 kHz and the relaxation time $T_1$ = 3.24 ms of the sample, we estimate theoretically a maximal contrast of -7.28% in the spin relaxometry experiment. Thus, principally, it is possible to realize the wide-field QNS with the modified setup. However, further optimization of the laser illumination and the signal readout is still necessary to increase the contrast.

In summary, we have built a camera-based integrated setup to demonstrate fully coherent control of NV center spin via Rabi oscillations, Ramsey, Hahn echo, and spin relaxation experiments, i.e. the core of the quantum noise spectroscopy experiments. In these experiments, we achieved a high contrast of 12% with a readout time of 500 μs, which is much longer than the typical readout time of hundreds of nanoseconds employed in the single NV center experiments. To understand the mechanism invovled, we have explored theoretically the NV center dynamics during these experiments, and explained the high contrast with the optical spin readout mechanism under extremely weak laser illumination, where spin-lattice relaxation and weak optical pumping result in a rather slow reduction of contrast with increasing integration time. By analyzing the limitations of the current setup, we have outlined the routes to upgrade the setup to carry out the widefield QNS experiments. The revealed readout mechanism and the upgraded device will pave the way to realize the characteristic time mapping of shallow NV centers in diamond, from which the magnetic field fluctuation from kHz to GHz can be extracted to analyze phase-transition of magnetic or superconducting materials.

## ACKNOWLEDGMENTS

Jiahui Chen carried out the studies under the supervision of Yuan Zhang, who designs the experiment and the theory. We thank Xun Yang for diamond sample preparation, Chong-Xin Shan for help with guiding opinions in the process of experimental and theoretical work, and other authors for help with discussions and assistance in experiments. This work was supported by the National Key R&D Program of China (Grant No. 2024YFE0105200), the National Natural Science Foundation of China (Grants No. 12422413 and No. 62475242), Science and Technology Major Project of Henan Province (231100230300), Henan Association for Science and Technology Youth Talent Support Program (2024HYTP024), Natural Science Foundation of Henan Province (252300421228), Central Plains Outstanding Young Talents Program.

# Supplemental Material for "Long-exposure Camera Readout for Coherent Control of Nitrogen-Vacancy Center Spins in Diamond"

Jiahui Chen[1†], Qilong Wu[1†], Huihui Yu[1], Yi-Dan Qu[1], Yuan Zhang[1,2*], Xun Yang[1*], Chong-Xin Shan[1*]

1. Henan Key Laboratory of Diamond Optoelectronic Materials and Devices, Key Laboratory of Material Physics Ministry of Education, School of Physics and Microelectronics, Zhengzhou University, Zhengzhou 450052, China
2. Institute of Quantum Materials and Physics, Henan Academy of Sciences, Zhengzhou 450046, China

†The authors contribute equally to the work.
*Contract emails: yzhuaudipc@zzu.edu.cn; yx@zzu.edu.cn; cxshan@zzu.edu.cn

## I. Details of Camera-based Integrated Setup

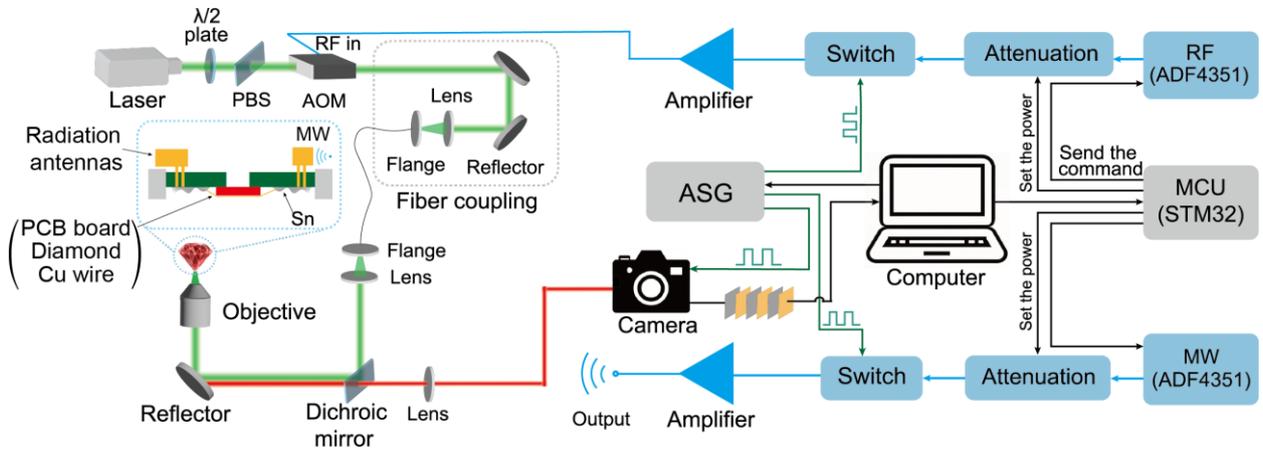

**FIG. S1. Schematic of the whole system including an optical module (left) and an electric module (right).** The optical module includes a laser system, an inverted optical microscope, where the sample holder is enlarged. The electronic module includes the electronics for the chip-based microwave and radiofrequency source, the power amplifier and a digital delay/pulse generator. For more details, see the text.

In this section, we describe the details of the camera-based integrated setup (Fig. S1). The left half shows the schematics of the optical module, which includes a laser system and an inverted optical microscope. In the laser system, a diode laser is powered by a constant current source, and provides a beam with a 532 nm wavelength and a maximal power up to 130 mW. The laser beam is attenuated firstly by a half-wave plate and polarization beam splitter, and reflected by a mirror. After passing through the AOM, the laser beam is further reflected by two mirrors, and finally coupled into a single-mode fiber. Using the radiofrequency radiation to power the AOM, the 1st-order diffraction beam of power 16.8 mW can be obtained from the incoming beam of 50 mW, leading to a diffraction efficiency of 33.6%. The low efficiency is currently limited by a relatively large olive-shaped beam out of the diode laser, and can be further optimized by collimating the diode with an aspheric lens. By adjusting another two mirrors, the power of the output beam can be increased up to 9.7 mW, leading to a coupling efficiency of 57.7%.

In the inverted optical microscope, the laser beam out of a single-mode fiber is delivered into the diamond sample after being reflected by a dichroic mirror (OFD1LP-650) and focused by an objective (SLMPlan 50×). The NV fluorescence is collected by the same objective in an inverted direction after being filtered by the dichroic mirror and a 650 nm long pass filter (OFE1LP-650), and is then detected by the industrial camera (Daheng MR2-160-249U3M-HS-6P). Two 3D translation stages are used to adjust the positions of the diamond sample and a permanent magnet. The diamond sample is mounted on a PCB board integrated with a microwave antenna. A 15 μm diameter copper wire is used to deliver the microwave radiation to the diamond sample. The laser system and the optical microscope are installed on two optical breadboards, which increases the flexibility significantly.

The right half shows the structure of the electronic module, which consists of a computer, a microcontroller (STM32F103ZET), a digital delay/pulse generator (ASG 8100 from CIQTEK), a microwave and a radiofrequency (RF) output module. The digital delay/pulse generator provides three channels of Transistor-Transistor Logic (TTL) pulses to switch on and off the RF radiation (and thus the laser beam through the AOM), the microwave radiation and to trigger the digital camera to take the fluorescence images of the diamond sample. The microcontroller is programmed to send the commands to two microwave source chips (ADF4351), and two digital RF attenuator chips (HMC624A). Two switchers (ZASWA-2-50DAR+) are used to switch on and off the microwave and radiofrequency radiation in tens of nanoseconds, and two amplifiers (KDT700MPA-035, KDT2038PA-045) amplify them by up to 36 dB and 30 dB, respectively.

A laptop is connected with the microcontroller, the digital camera and the pulse generator via USB cables. Several python programs are written to communicate with these devices. To organize the flow of the experiments, we have prepared the programs with Jupyter notebook scripts, which provide not only necessary information but also allow to structure the codes with blocks. We have updated the firmware to the microcontroller so that it can understand the commands delivering through serial port. We have relied on the functions in the third-party library offered by DaHeng and CIQTEK to control the camera and the pulse generator.

## II. Quantum Master Equation and Python Code

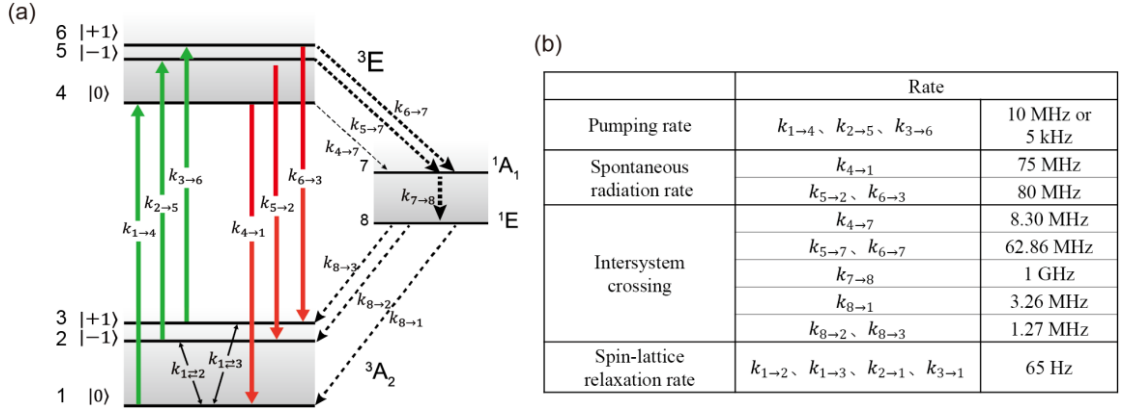

**FIG. S2. Labeling of the NV center levels (a) and the summary of the rates for various processes (b).** For more details, see the text.

In Sec. III of the main text, we have compared theoretically the readout mechanism of the NV center under the strong and weak laser illumination. For this theoretical study, we label the eight levels of the NV centers as $^3A_2, m_s = 0, -1, +1 \rightarrow i = 1,2,3$, $^3E$, $m_s = 0, -1, +1 \rightarrow i = 4,5,6$, and $^1A_1, ^1E \rightarrow i = 7,8$ [Fig. S2(a)], and then establish a quantum master equation for reduced density operator $\hat{\rho}$:

$$\frac{\partial}{\partial t}\hat{\rho} = -\sum_{i,j} k_{i \rightarrow j}\left[\frac{1}{2}\{\hat{\sigma}_{ij}\hat{\sigma}_{ji}, \hat{\rho}\} - \hat{\sigma}_{ji}\hat{\rho}\hat{\sigma}_{ij}\right] \tag{S1}$$

to describe the various dynamics with the operators $\hat{\sigma}_{ij}$ and rates $k_{i \rightarrow j}$ [Fig. S2(b)]. The values of $k_{i \rightarrow j}$ are taken from those for single NV center under the diffraction-limited laser illumination. In particular, the rates $k_{1 \rightarrow 4} = k_{2 \rightarrow 5} = k_{3 \rightarrow 6} = \eta = 30$ MHz/mW describe the optical pumping from the $^3A_2$, m$_s$ = 0,+1,-1 levels to the $^3E$, m$_s$ = 0,+1,-1 levels. We solve the master equation with QuTip[2], and then extract the level population $P_i$, and the fluorescence intensity $PL = k_{4 \rightarrow 1}P_4 + k_{5 \rightarrow 2}P_5 + k_{6 \rightarrow 3}P_6$.

```
(a)
1.  import matplotlib.pyplot as plt
2.  import numpy as np
3.  from matplotlib.ticker import MultipleLocator
4.  from qutip import *
5.  l1 = basis(8,0); l2 = basis(8,1); l3 = basis(8,2); l4 = basis(8,3);
6.  l5 = basis(8,4); l6 = basis(8,5); l7 = basis(8,6);l8 = basis(8,7);
7.  eta_0 = 5e3; k14 = eta_0; k25 = eta_0; k36 = eta_0;
8.  k41 = 0.075e9 ; k52 = 0.08e9; k63 = 0.08e9;
9.  k47 = 0.0083e9; k57 = 0.062857e9; k67 = 0.062857e9;
10. k78 = 1e9 ;
11. k81 = 0.0032558e9; k82 = 0.001279e9; k83 = 0.001279e9;
12. gamma = 65 ;
13. k12 = gamma; k13 = gamma;
14. k21 = gamma; k31 = gamma;
15. H = 0*l1*l1.dag() ;
16. psi0 = 1/3*l1*l1.dag() + 1/3*l2*l2.dag() + 1/3*l3*l3.dag() ;
17. c_op = [np.sqrt(k14)*l4*l1.dag(),np.sqrt(k25)*l5*l2.dag(),
         np.sqrt(k36)*l6*l3.dag(), np.sqrt(k41)*l1*l4.dag(),
         np.sqrt(k52)*l2*l5.dag(), np.sqrt(k63)*l3*l6.dag(),
         np.sqrt(k47)*l7*l4.dag(), np.sqrt(k57)*l7*l5.dag(),
         np.sqrt(k67)*l7*l6.dag(), np.sqrt(k78)*l8*l7.dag(),
         np.sqrt(k81)*l1*l8.dag(),np.sqrt(k82)*l2*l8.dag(),
         np.sqrt(k83)*l3*l8.dag(), np.sqrt(k12)*l2*l1.dag(),
         np.sqrt(k13)*l3*l1.dag(), np.sqrt(k21)*l1*l2.dag(),
         np.sqrt(k31)*l1*l3.dag(), np.sqrt(k45)*l5*l4.dag(),
         np.sqrt(k46)*l6*l4.dag(),np.sqrt(k54)*l4*l5.dag(),
         np.sqrt(k64)*l4*l6.dag()]
18. tlist_0 = np.linspace(0,3200e-6,10000)
19. me0 = mesolve(H, psi0, tlist_0, c_op)
20. p1_0, p2_0, p3_0, p4_0, p5_0, p6_0, p7_0, p8_0 = [expect
    (l * l.dag(), me0.states) for l in [l1, l2, l3, l4, l5, l6, l7, l8]]
21. eta = 0.0; k14 = eta; k25 = eta; k36 = eta
22. psi1 = me0.states[-1]

(b)
23. H = 0*l1*l1.dag()
24. tlist_1 = np.linspace(0,1e-6,10000)
25. c_op = [np.sqrt(k14)*l4*l1.dag(),np.sqrt(k25)*l5*l2.dag(),
         np.sqrt(k36)*l6*l3.dag(), np.sqrt(k41)*l1*l4.dag(),
         np.sqrt(k52)*l2*l5.dag(), np.sqrt(k63)*l3*l6.dag(),
         np.sqrt(k47)*l7*l4.dag(), np.sqrt(k57)*l7*l5.dag(),
         np.sqrt(k67)*l7*l6.dag(), np.sqrt(k78)*l8*l7.dag(),
         np.sqrt(k81)*l1*l8.dag(),np.sqrt(k82)*l2*l8.dag(),
         np.sqrt(k83)*l3*l8.dag(), np.sqrt(k12)*l2*l1.dag(),
         np.sqrt(k13)*l3*l1.dag(), np.sqrt(k21)*l1*l2.dag(),
         np.sqrt(k31)*l1*l3.dag(), np.sqrt(k45)*l5*l4.dag(),
         np.sqrt(k46)*l6*l4.dag(),np.sqrt(k54)*l4*l5.dag(),
         np.sqrt(k64)*l4*l6.dag()]
26. me1 = mesolve(H, psi1, tlist_1, c_op)
27. p1_1, p2_1, p3_1, p4_1, p5_1, p6_1, p7_1, p8_1 = [expect
    (l * l.dag(), me1.states) for l in [l1, l2, l3, l4, l5, l6, l7, l8]]
28. eta = eta_0
29. k14 = eta_0; k25 = eta_0; k36 = eta_0
30. psi2 = me1.states[-1]
31. H = 0*l1*l1.dag()
32. tlist_2 = tlist_0
33. c_op = [np.sqrt(k14)*l4*l1.dag(),np.sqrt(k25)*l5*l2.dag(),
         np.sqrt(k36)*l6*l3.dag(), np.sqrt(k41)*l1*l4.dag(),
         np.sqrt(k52)*l2*l5.dag(), np.sqrt(k63)*l3*l6.dag(),
         np.sqrt(k47)*l7*l4.dag(), np.sqrt(k57)*l7*l5.dag(),
         np.sqrt(k67)*l7*l6.dag(), np.sqrt(k78)*l8*l7.dag(),
         np.sqrt(k81)*l1*l8.dag(),np.sqrt(k82)*l2*l8.dag(),
         np.sqrt(k83)*l3*l8.dag(), np.sqrt(k12)*l2*l1.dag(),
         np.sqrt(k13)*l3*l1.dag(), np.sqrt(k21)*l1*l2.dag(),
         np.sqrt(k31)*l1*l3.dag(), np.sqrt(k45)*l5*l4.dag(),
         np.sqrt(k46)*l6*l4.dag(),np.sqrt(k54)*l4*l5.dag(),
         np.sqrt(k64)*l4*l6.dag()]

(c)
34. me2 = mesolve(H, psi2, tlist_2, c_op)
35. p1_2, p2_2, p3_2, p4_2, p5_2, p6_2, p7_2, p8_2 = [expect
    (l * l.dag(), me2.states) for l in [l1, l2, l3, l4, l5, l6, l7, l8]]
36. fig, axs = plt.subplots(1, 3, figsize=(12, 6), sharey=True)
37. for p in [p1_0, p2_0, p3_0, p4_0, p5_0, p6_0, p7_0, p8_0]:
        axs[0].plot(tlist_0 / 1e-3, p, linewidth=1)
38. for p in [p1_1, p2_1, p3_1, p4_1, p5_1, p6_1, p7_1, p8_1]:
        axs[1].plot(tlist_1 / 1e-6, p, linewidth=1)
39. for p in [p1_2, p2_2, p3_2, p4_2, p5_2, p6_2, p7_2, p8_2]:
        axs[2].plot(tlist_2 / 1e-3, p, linewidth=1)
40. plt.show()
41. PL0 = p4_0*k41 + p5_0*k52 + p6_0*k63
42. PL1 = p4_1*k41 + p5_1*k52 + p6_1*k63
43. PL2 = p4_2*k41 + p5_2*k52 + p6_2*k63
44. IPL0 = np.zeros(PL0.shape[0])
45. for ind in range(0,PL0.shape[0]):
        IPL0[ind] = np.sum(PL0[0:ind])
46. IPL1 = np.zeros(PL1.shape[0])
47. for ind in range(1,PL1.shape[0]):
        IPL1[ind] = np.sum(PL1[0:ind])
48. IPL2 = np.zeros(PL2.shape[0])
49. for ind in range(1,PL2.shape[0]):
        IPL2[ind] = np.sum(PL2[0:ind])
50. 
51. contrast = (IPL0[1:]-IPL2[1:])/IPL2[1:]
52. plt.plot(tlist_0/1e-6, PL0/1e6)
53. plt.plot(tlist_2/1e-6, PL2/1e6)
54. plt.twinx().plot(tlist_0[2:]/1e-6, contrast[1:]*100)
55. plt.show()
```

**FIG. S3. Program codes to solve the quantum master equation.** For more details, see the text.

In Fig. S3, we present the Python codes to solve the master equation (S1) using QuTip package. Lines 1st - 4th in Fig. S4(a) import the necessary packages. Lines 5th - 6th define the NV centers as the eight-level systems. Lines 7th - 14th define the optical pumping rate, spontaneous emission rate, intersystem crossing rate, and spin-lattice relaxation rate. Lines 15th - 27th describe the initial state of the system. Lines 15th - 16th describe the zero Hamiltonian, and the initial reduced density operator, resulting equal population on the spin levels $^3A_2, m_s = 0, -1, +1$. Line 17th defines the list of operators, which are used later to define the Lindblad terms. Line 18th defines the list of times. Line 19th calculates the steady state of the system using parameters defined above. Line 20th extracts the steady-state population of the eight-levels. Lines 21st - 27th describe the free evolution process without laser pumping. The code structure is similar to the previous section, except that line 21st sets the pumping rate as zero, and line 22nd uses the steady state reduced density operator from the preceding stage as the initial state for the current simulation. Likewise, lines 28th - 35th describe the readout process. Lines 36th - 40th plot the population of the eight energy levels during the three stages. Lines 41st - 43rd calculate the fluorescence intensity for each stage. Lines 44th - 47th calculate the fluorescence for different integration times, and lines 48th – 55th compute the contrast. Finally, lines 52nd - 55th plot the dynamics of the signal, reference and contrast, respectively.

# III. Effective Three Levels Model of NV Center under Weak Laser Illumination

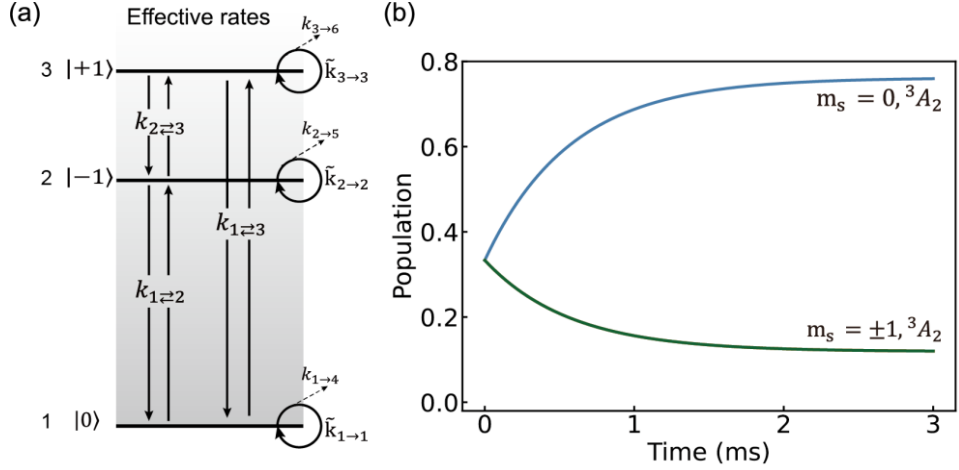

**FIG. S4. Effective three-level model.** (a) shows the energy diagram of the NV center under weak laser pumping after eliminating adiabatically the excited levels. (b) shows the calculated population dynamics of the $^3A_2$, $m_s=\pm1$ levels, which is comparable with Fig. 3(c).

In the main text, we show that the excited levels are almost unpopulated for the NV centers under the weak laser illumination, and the population is transferred effectively between the $^3A_2$, $m_s=0$ level and the $^3A_2$, $m_s=\pm1$ level [Fig. 3 (c,d)]. Motivated by this observation, in this section, we eliminate adiabatically the excited levels in our theoretical model to achieve an effective three-levels model for the NV center under weak laser illumination. When using the QuTip package to solve the master equation (S1), one essentially represents the density operator $\hat{\rho}$ as a matrix with elements $\rho_{ij}$ in the Hilbert space formed by the eight NV levels, and then derives the corresponding equations for these elements. Since there is no coherent process involved, only the diagonal elements $\rho_{ii}$ are non-zero, and represent the populations on the corresponding levels $P_i$. These populations satisfy essentially the following rate equations:

$$\partial_t P_1 = -(k_{1\to 2} + k_{1\to 3} + k_{1\to 4})P_1 + \sum_{i=4,8,2,3} k_{i\to 1} P_i \tag{S2}$$

$$\partial_t P_2 = -(k_{2\to 1} + k_{2\to 5})P_2 + \sum_{i=5,8,1} k_{i\to 2} P_i \tag{S3}$$

$$\partial_t P_3 = -(k_{3\to 1} + k_{3\to 6})P_2 + \sum_{i=6,8,1} k_{i\to 3} P_i \tag{S4}$$

$$\partial_t P_4 = -(k_{4\to 1} + k_{4\to 7})P_4 + k_{1\to 4} P_1 \tag{S5}$$

$$\partial_t P_5 = -(k_{5\to 2} + k_{5\to 7})P_5 + k_{2\to 5} P_2 \tag{S6}$$

$$\partial_t P_6 = -(k_{6\to 3} + k_{6\to 7})P_6 + k_{3\to 6} P_3 \tag{S7}$$

$$\partial_t P_7 = -k_{7\to 8} P_7 + \sum_{i=4,5,6} k_{i\to 7} P_i \tag{S8}$$

$$\partial_t P_8 = -(k_{8\to 1} + k_{8\to 2} + k_{8\to 3})P_8 + k_{7\to 8} P_7 \tag{S9}$$

Considering Eqs. (S4-S8) at the steady-state, we express the population of the excited levels $P_i$ ($i = 4, ..., 8$) with that of the ground levels $P_i$ ($i = 1, ..., 3$). Inserting these expressions into Eqs. (S2-S3), we obtain

$$\partial_t P_1 = -(k_{1\to 3} + k_{1\to 2} + k_{1\to 4} - \tilde{k}_{1\to 1})P_1 + (\tilde{k}_{1\to 2} + k_{2\to 1})P_2 + (\tilde{k}_{1\to 3} + k_{3\to 1})P_3 \tag{S10}$$

$$\partial t P_2 = -(k_{2\to 1} + k_{2\to 5} - \tilde{k}_{2\to 2})P_2 + (\tilde{k}_{2\to 1} + k_{1\to 2})P_1 + \tilde{k}_{2\to 3} P_3 \tag{S11}$$

$$\partial t P_3 = -(k_{3\to 1} + k_{3\to 6} - \tilde{k}_{3\to 3})P_3 + (\tilde{k}_{3\to 1} + k_{1\to 3})P_1 + \tilde{k}_{3\to 2} P_2 \tag{S12}$$

In the above equations, we have defined the following rates

$$\tilde{k}_{i \to i} = \frac{k_{i \to i+3}}{k_{i+3 \to i} + k_{i+3 \to 7}} \left[ k_{i+3 \to i} + \frac{k_{8 \to i} \cdot k_{i+3 \to 7}}{(k_{81} + k_{82} + k_{83})} \right] \tag{S13}$$

$$\tilde{k}_{i \to j} = \frac{k_{i+3 \to 7} \cdot k_{j \to j+3}}{(k_{j+3 \to j} + k_{j+3 \to 7})} \cdot \frac{k_{8 \to i}}{(k_{8 \to 1} + k_{8 \to 2} + k_{8 \to 3})}, (i,j = 1,2,3, i \neq j) \tag{S14}$$

In these equations, $\tilde{k}_{i \to i}$ (i=1,2,3) describe the population transferred to the excited levels, $\tilde{k}_{i \to j}$ ($i,j = 1,2,3, i \neq j$) describe the effective population transfer from the i-th spin level to the j-th spin level of the triplet ground state [Fig. S4(a)]. Fig. S5(b) shows the population dynamics for the NV center illuminated by a laser of 6 mW, which agrees perfectly with Fig. 3(c) and thus verifies the effective model developed here.

## IV. Extra Data

| | Second ODMR in Fig. 2 (b) | | | |
|---|---|---|---|---|
| $f_i$(MHz) | 2759.08 | 2841.59 | 2911.40 | 2981.12 |
| $\gamma_i$(MHz) | 15.48 | 17.41 | 15.48 | 14.04 |
| $C_i$ (%) | -2.11 | -4.38 | -4.27 | -2.12 |

**TAB. S1.** Parameters extracted from the CW-ODMR experiment.

Tab. S1 shows the parameters about the resonance frequency $f_i$, full width at half maximum $\gamma_i$, and fluorescence contrast $C_i$ extracted from the CW-ODMR experiment.